\documentclass[letterpaper,10pt,twocolumn]{article}
\usepackage{titling}
\usepackage[T1]{fontenc}
\usepackage{parskip}
\usepackage[font={small}]{caption}
\usepackage[hidelinks]{hyperref}
\usepackage[margin=2cm]{geometry}
\usepackage{times}
\usepackage{amsmath,amssymb,amsthm,amsfonts}
\usepackage[affil-it]{authblk}
\usepackage{listings}

\usepackage{graphicx}      
\usepackage{natbib}        
\usepackage{color}
\usepackage{bm}
\usepackage{listings}
\usepackage{subcaption}
\usepackage{tikz}
\usepackage{algorithm}
\usepackage[noend]{algpseudocode}
\usetikzlibrary{positioning}

\usepackage{pgfplots}
\pgfplotsset{compat=newest}

\usepackage{mymathsty}

\newcommand*{\thetabhls}{{\widehat{\thetab}_{\text{LS}}}}

\title{Data-Driven Impulse Response Regularization via Deep Learning }
\author[]{Carl Andersson} 
\author[]{Niklas Wahlstr\"{o}m} 
\author[]{Thomas B. Sch\"{o}n}
\date{}

\affil[]{Department of Information Technology, Uppsala University, Sweden. \\ Email:  \{carl.andersson, niklas.wahlstrom, thomas.schon\}@it.uu.se}

\begin{document}

\newsavebox{\citethis}
\begin{lrbox}{\citethis}
\begin{lstlisting}[breaklines,basicstyle=\small\ttfamily]
@inproceedings{CarlAndersson2018,
author    = {Carl Andersson and Niklas Wahlstr\"{o}m and Thomas Bo Sch\"{o}n},
title     = {Data-Driven Impulse Response Regularization via Deep Learning},
booktitle = {Proceedings of 18th IFAC Symposium on System Identification (SYSID)},
address   = {Stockholm, Sweden},
pages     = {1-6},
year      = {2018}
}
\end{lstlisting}
\end{lrbox}

\twocolumn[
\begin{@twocolumnfalse}
	\maketitle

	\textbf{Please cite this version:}

Carl Andersson, Niklas Wahlstr\"{o}m, Thomas B. Sch\"{o}n.   \textbf{Data-Driven Impulse Response Regularization via Deep Learning}. In \textit{Proceedings of 18th IFAC Symposium on System Identification (SYSID)}, Stockholm, Sweden,  2018.

\begin{center}
\begin{minipage}{.75\linewidth}
	\usebox{\citethis}
\end{minipage}
\end{center}
	
	\vspace{3em}
	
	\begin{abstract}  
	We consider the problem of impulse response estimation of stable linear single-input single-output systems. It is a well-studied problem where flexible non-parametric models recently offered a leap in performance compared to the classical finite-dimensional model structures. Inspired by this development and the success of deep learning we propose a new flexible data-driven model. Our experiments indicate that the new model is capable of exploiting even more of the hidden patterns that are present in the input-output data as compared to the non-parametric models.
\end{abstract}
\vspace{3em}

\end{@twocolumnfalse}
]	
 
\maketitle
	\begin{abstract}  
	We consider the problem of impulse response estimation of stable linear single-input single-output systems. It is a well-studied problem where flexible non-parametric models recently offered a leap in performance compared to the classical finite-dimensional model structures. Inspired by this development and the success of deep learning we propose a new flexible data-driven model. Our experiments indicate that the new model is capable of exploiting even more of the hidden patterns that are present in the input-output data as compared to the non-parametric models.
\end{abstract}

\section{Introduction}

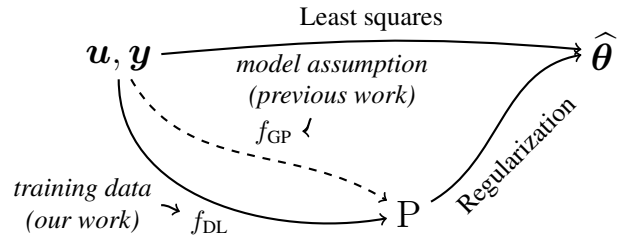
\begin{figure}[t]
\center
\begin{tikzpicture}
	[align=center,node distance=3cm]

	\node[] (x1) {\Large $\bm{u},\bm{y}$};
	\node[below right=1.5cm and 3cm  of x1] (x2) {\Large $P$};
	\node[above right=1.5cm and 2cm  of x2] (x3) {\Large $\thetabh$};

	\draw[dashed,->,thick] (x1) to [out=-60,in=148] node [above right,name=fgp] {$f_{\text{GP}}$} (x2) ;	
	\draw[->,thick] (x1) to [out=-90,in=-170] node [below,name=fdl] {$f_{\text{DL}}$} (x2);
	
	\draw[->,thick] (x2) to [out=30,in=-170] node [below=0.3cm, rotate=45] {Regularization} (x3);
	\draw[->,thick] (x1) to [out=5,in=175] node [above] {Least squares} (x3);
	
	\node[above right=-0.1cm and -1.0cm of fgp ]  (fgpn) {\textit{model assumption }\\ \textit{(previous work)}};

	\node[below left=-1.0cm and 0.25cm of fdl]  (fdln) {\textit{training data}\\  \textit{(our work)}};
 
 	\draw[->,thick] (fgpn) to [out=-120,in=2]  (fgp);
 	\draw[->,thick] (fdln) to [out= -0,in=-200]  (fdl);

\end{tikzpicture}
\caption{Schematic figure over the proposed method for impulse response estimation using deep learning in relation to the previous work using Gaussian processes. The functions $f_\text{GP}$ and $f_\text{DL}$ maps the input sequence $\bm{u}$ and the output sequence $\bm{y}$ of a system to an inverse regularization matrix $P$ for the Gaussian process approach and the deep learning approach, respectively.}
\label{fig:concept}
\end{figure}

Impulse response estimation has for a long time been at the core of system identification. Up until some five to seven years ago, the generally held belief in the field was indeed that we knew all there was to know about this topic. However, the enlightening work by \citet{Pillonetto2010} changed this by showing that the estimate can in fact be improved significantly by placing a Gaussian Process (GP) prior on the impulse response, which acts as a regularizer. This model-driven approach has since then been further refined \citep{Pillonetto2011,Chen2012,PillonettoDCNL:2014}, where the prior in this case could be interpreted to encode not only smoothness information, but also information about the exponential decay of the impulse response. In this paper we employ deep leaning (DL) to find a suitable regularizer via a method that is driven by data. Fig. \ref{fig:concept} depicts the general idea and the similarity of our method compared to the method based on Gaussian processes.

Deep learning is a  fairly new area of research that continues the work on neural networks from the 1990's. To get a brief, but informative, overview of the field of deep learning we recommend the paper by \citet{LeCun2015} and for a more complete snapshot of the field we refer to the monograph by \citet{GoodfellowBC:2016}. Deep learning has recently revolutionized several fields, including image recognition (e.g. \citet{Cirean2011}) and speech recognition (e.g. \citet{Hinton2012}). In both fields, deep learning has surpassed domain specific methods and hand-crafted feature design, by making use of large quantities of data in order to learn data-driven neural network models as general function approximators. 

The idea of using neural networks within system identification is certainly not new and they have been a standard tool for a long time, see e.g. \citet{SjobergZLBDGHJ:1995}. However, the current development in deep learning is  different from the past due to advances in computational software and hardware. As a consequence, contemporary neural networks are more reproducible than before which has increased the credibility of the area. Finally, the amount of available data has sky-rocketed which has made it possible to train larger models with better results. We believe that system identification still has lots to gain from using deep learning and this paper is just one concrete example of what can be done.

\section{Background \& problem formulation}
Consider a stable single-input single-output time-invariant linear system~$G^0$ relating an input sequence~$u(t)$ to an output sequence~$y(t)$ according to
\begin{align} 
	\label{eq:system}
	y(t) = G^0(q) u(t) + v(t),
\end{align}
where $q$ denotes the shift operator $qu(t) = u(t+1)$ and $v(t)$ denotes additive noise. The noise is assumed to be Gaussian white noise with zero mean and variance~$\sigma^2$. The system~$G^0(q)$ is represented using a  transfer function
\begin{align}
	G^0(q) = \sum_{k=1}^\infty g_k^0 q^{-k},
\end{align}
where $g_1^0, g_2^0, \dots$ denote the impulse response coefficients of the system. We make use of superscript~$0$ to denote the true system.

Based on data, i.e. an input sequence $\bm{u} = (u(1),\dots,u(N))^T $ and an output sequence $\bm{y} = (y(1),\dots,y(N))^T $ both of length~$N$, the task is to compute an estimate that represents the true system~$G^0(q)$ as well as possible. 

Traditionally, the transfer function is encoded using a finite number of parameters~$\thetab$ ($\text{dim}(\thetab) = n \ll N$), for example, via a finite impulse response (FIR) model,
\begin{align} \label{eq:fir}
	 G(q; \thetab) = B(q) = b_1 q^{-1} + \dots + b_{n_b} q^{-n_b},
\end{align}	 
or an output-error (OE) model,
\begin{align}
G(q; \thetab) = \frac{B(q)}{F(q)} = \frac{b_1 q^{-1} + \dots + b_{n_b} q^{-n_b}}{f_1 q^{-1} + \dots + f_{n_f} q^{-n_f}},
\end{align}
where $\thetab =  (b_1,\dots,b_{n_b}  )^\T$ or $\thetab =  ( b_1,\dots,b_{n_b},f_1,\dots,f_{n_f} )^\T$, respectively.
See \citet{Ljung1999} for a comprehensive list of model structures. With these model structures, the prediction error method can be used to compute a point estimate $\thetabh$ of the unknown parameters by solving the following optimization problem,
\begin{align} \label{eq:pem_reg}
	\thetabh = \argmin_\thetab \sum_{t=1}^N (y(t) - G(q, \thetab)u(t))^2 + \thetab^\T D \thetab,
\end{align}
where $ \thetab^\T D \thetab$ describes the added regularization term, governed by the regularization matrix~$D$. It has been shown by \citet{Pillonetto2010,Pillonetto2011,Chen2012} that the use of an effective regularization is more important than the specific choice of model-order,~$n$. Intuitively the use of a GP model and the regularization term $ \thetab^\T D \thetab$ opens up for model selection at a "finer" scale compared to what is possible with the classical finitely parameterized model structures. Adding this kind of regularization is especially important when the number of samples, $N$, is low compared to the parameters we wish to estimate, $n$. 

In the case of the FIR model, the optimization problem reduces to a linear least squares problem to find the optimal parameters~$\thetabh$. For numerical reasons the inverse regularization matrix~$P=D^{-1}$ is often used in place of~$D$. This gives rise to the following analytic solution,
\begin{subequations}
	\label{eq:LS_reg}
	\begin{align}
		\thetabh =  \left (PR+I_{n}\right ) ^{-1} P F_N,
	\end{align}
	where 
	\begin{align}
		R &= \sum_{t = n+1}^N \varphi(t) \varphi(t)^\T, \\
		\varphi(t) &=  \lbrace u(t-1) \dots u(t-n)  \rbrace^\T, \\
		F_N &= \sum_{t = n+1}^N \varphi(t) y(t).
	\end{align}
\end{subequations}
We will throughout this paper denote then estimate~\eqref{eq:LS_reg} by $\widehat{\thetab}(P)$ to stress its dependence on the inverse regularization matrix~$P$. As a special case, with $P = 0$, we have the least squares solution, which we denote $\thetabhls$.

One question still remains though; how do we find the inverse regularization matrix~$P$? One of the most general ideas is to let it  depend on $\bm{u}$ and $\bm{y}$. This was done in \citet{Pillonetto2010,Pillonetto2011,Chen2012} by modelling the impulse response as a Gaussian process. A Gaussian process is known to be a very flexible prior, even so, since the model only depends on a low number of hyper-parameters, typically one for a lengthscale and one for some noise, it heavily depends on the specific model we choose for the covariance function. These hyper-parameters are replaced by a point estimate obtained by maximizing the marginal likelihood of the observed data, a procedure known as Empirical Bayes \citep{Bishop:2006}. The regularization matrix will thus implicitly depend on $\bm{u}$ and $\bm{y}$ via the hyper-parameters. We explicitly denote this dependence by~$P=f_\text{GP}(\bm{u},\bm{y})$. This method is also explained in more detail in Section~\ref{sec:Reg_GP}.

We instead propose an arguably even more flexible model by parametrizing the inverse regularization matrix~$P$ with a neural network $P = f_\text{DL}(\bm{u},\bm{y})$. In contrast to the Gaussian process model these parameters are computed by training the model with lots of training data consisting of an input sequence, an output sequence and the true impulse response for either real or simulated systems. Compared to the GP model this is a more data-driven approach to the problem which also makes it possible to use existing techniques from deep learning when building and training the model.

\section{Regularization using \\ Gaussian Process}
\label{sec:Reg_GP}
The Gaussian prior offers a natural way of encoding the smoothness and decay characteristics that we find in the impulse response from a stable linear system. The specific details of these characteristics are tuned via the hyper-parameters~$\lambda$. The resulting GP prior can be written as
\begin{align}
	\thetab \sim p_\lambda(\thetab) =  \mathcal{N}(\thetab \given 0, P^{\lambda}),
\end{align}
where $P^\lambda$ in this Bayesian setting is equal to the inverse regularization matrix in~\eqref{eq:LS_reg}. The matrix~$P^\lambda$ is related to the covariance function of the Gaussian Process as~$k_{{\lambda}} (i, j)=~P_{ij}^{\lambda}$,
where $P_{ij}^{\lambda}$ denotes then entry on row~$i$ and column~$j$ of~$P^{\lambda}$. With the aid of measured data we can select a point estimate of the hyper-parameters~$\widehat{\lambda}$ by maximizing the marginal log-likelihood,
\begin{align}
\label{eq:EB}
	\widehat{\lambda} = \argmax_\lambda \log \int  p(\bm{y} \given \thetab;\bm{u}) p_\lambda(\thetab) d\thetab,
\end{align}
where 
\begin{align}
	p(\bm{y} \given \thetab;\bm{u})  = \prod_{t=1}^N p(y(t) \given \thetab;\bm{u})  =   \prod_{t=1}^N \mathcal{N}(y(t) \given\thetab^\T \varphi(t),\sigma^2).
\end{align}
As a direct consequence of this, the optimal inverse regularization matrix~$P^{\widehat\lambda}$ implicitly depends on the input sequence~$\bm{u}$ and the output sequence~$\bm{y}$. The equation \eqref{eq:EB} then describe the function $P = f_\text{GP}(\bm{u},\bm{y})$. Using this inverse regularization matrix together  with~\eqref{eq:LS_reg}, we obtain an estimate of the FIR parameters that has better accuracy and is more robust than the non-regularized approach~\citep{Chen2012}.

\section{Regularizing using Deep Learning}
\label{sec:DL}
In contrast to previous work where the regularization matrix only depends on a few hyper-parameters we instead model the regularization matrix directly with a neural network that depends on a large number of parameters~$\eta$  according to
\begin{align}
	\label{eq:P_DL}
	P = f_\text{DL}(\bm{u},\bm{y};\eta). 
\end{align}
To select specific values for all these parameters we start by formulating the mean squared error (MSE) of the estimate~$\thetabh(P)$ in~\eqref{eq:LS_reg}, the so-called estimation error,
\begin{align}
\label{eq:MSE}
	\text{MSE}\left (\thetabh \right ) = \left \lVert \thetabh( f_\text{DL}(\bm{u},\bm{y};\eta) ) - \thetab_0  \right \rVert^2,
\end{align}
where $\lVert \cdot \rVert^2$, denotes the 2-norm and~$\thetab_0$ denotes the true FIR parameters which corresponds to the truncated true impulse response. Here we make use of the neural network model~\eqref{eq:P_DL} of~$P$ when forming the MSE. The parameters~$\widehat \eta$ to use in~\eqref{eq:P_DL} are found by simply minimizing  this estimation error,
\begin{align} \label{eq:est_sum}
	 \widehat{\eta} = \argmin_\eta \frac{1}{M} \sum_{i=1}^M  \left \lVert \thetabh \left (f_\text{DL}(\bm{u}^{(i)},\bm{y}^{(i)};\eta) \right ) - \thetab_0^{(i)}\right \rVert^2,
\end{align} 
where $M$ is the number of training examples. To set the terminology straight we use the term \emph{training} for the minimization of the estimation error w.r.t.~$\eta$ whereas \emph{estimation} refers to the computation of a point estimate of the FIR parameters by minimizing the one step prediction error w.r.t.~$\thetab$.

The following sections describe our method and an overview is provided in Algorithm \ref{alg:algorithm}.

\subsection{Regularization model}
\label{sec:RegMod}
We know that~$P$ must be positive  semi-definite. Inspired by this fact we choose~$P$ as a weighted sum of rank-one positive definite matrices. The idea behind this choice is to have a representation that is flexible enough to represent all possible regularization matrices and at the same time encode the knowledge that the optimal regularization matrix is a rank-one matrix (see Appendix~\ref{sec:opt_reg}).

Our rank-one matrices are constructed as outer products of a vector $\bm{s}_i$ with itself where the elements of the vectors are free parameters. The weights, $w_i$, that weight our rank-one matrices, are modelled as the output of a softmax layer from a neural network. The input to this neural network is $\bm{u}$ and $\bm{y}$ as well as the nonregularized least squares solution, which only depends on $\bm{u}$ and $\bm{y}$. Hence, we have,
\begin{align} \label{eq:reg_mat_dl}
	 P  = \sum_{i=1}^{n_m} w_i(\bm{u},\bm{y},\thetabhls;{\eta'}) \bm{s}_i \bm{s}_i^\T,
\end{align}
where $n_m$ is the number of rank-one matrices used to represent the regularization matrix and $\eta'$ is the parameters of the neural network. We use $\eta = \lbrace \eta',\bm{s}_1,\dots,\bm{s}_{n_m} \rbrace$ to collect all the parameters in the regularization matrix model.

\subsection{Neural network model}
\label{sec:NetworkMod}
We have made use of four fully connected layers in our neural network, where the final layer is a softmax layer producing  weights between~$0$ and~$1$. 
The other activation functions for the fully connected layers are Rectified Linear Units (ReLU) which are defined as $\text{ReLU}(x) = \max(0,x)$. A typical layer of the network is thus written as 

$\bm{h}^{(i+1)} = \text{ReLU}(W \bm{h}^{(i)} + \bm{b})$, 
where~$\bm{h}^{(i)}$ denotes the input to the layer, $\bm{h}^{(i+1)}$ denotes the output from the layer, $W$ denotes a so-called weight matrix of dimensions~${\text{dim}(\bm{h}^{(i+1)})\times\text{dim}(\bm{h}^{(i)})}$ 
and~$\bm{b}$ denotes a so-called bias term of dimension~$\text{dim}(\bm{h}^{(i)})$. Both the weight matrices and the bias terms are part of the parameters~$\eta'$ describing the network, i.e.~$W^{(i)} \in \eta'$ and~$\bm{b}^{(i)} \in \eta'$. To regularize the training procedure we add a dropout layer after the softmax layer which with~$30\%$ probability sets a  weight to zero during training. This is a standard technique to avoid overfitting in NN \citep{Srivastava2014}. The input to the network is the input~$\bm{u}$ and output~$\bm{y}$ sequence of the system we want to estimate and the corresponding non-regularized least squares solution~$\thetabhls$. 

\begin{figure}[ht]
\center
\begin{tikzpicture}
	[align=center,node distance=1.2cm, minimum size=0.8cm]

	\node[rectangle,thick,draw] (h1) {$\bm{h}^{(1)} = \text{ReLU}\left (W^{(1)}\{\bm{u},\bm{y},\thetabhls\} + \bm{b}^{(1)} \right )$};

	\node[circle,thick,draw,above of=h1,left of=h1] (x1) {$\bm{u}$};
	\node[circle,thick,draw,above of=h1] (x2) {$\bm{y}$};
	\node[circle,thick,draw,above of=h1,right of=h1] (x3) {$\thetabhls$};
	
	\node[rectangle,thick,draw,below of=h1] (h2) {$\bm{h}^{(2)} = \text{ReLU}\left (W^{(2)}\ \bm{h}^{(1)} + \bm{b}^{(2)} \right )$};
	
	\node[rectangle,thick,draw,below of=h2] (h3) {$\bm{h}^{(3)} = \text{ReLU}\left (W^{(3)}\ \bm{h}^{(2)} + \bm{b}^{(3)} \right )$};

	\node[rectangle,thick,draw,below of=h3,minimum width=4cm] (h4) {$\bm{h}^{(4)} = W^{(4)}\ \bm{h}^{(3)} + \bm{b}^{(4)}$};
	
	\node[rectangle,thick,draw,below of=h4,minimum width=4cm] (h5) {$\bm{w}_d = \frac{\text{exp}\left( \bm{h}^{(4)} \right)}{1+\sum_{i=1}^{n_m} \text{exp}\left(h^{(4)}_i \right)}$};

	\node[rectangle,thick,draw,below of=h5,minimum width=4cm] (h6) {$\bm{w} = \text{dropout}(\bm{w}_d)$};

	\draw[->,thick] (x1) edge (h1.north -| x1);
	\draw[->,thick] (x2) edge (h1);
	\draw[->,thick] (x3) edge (h1.north -| x3);
	
	\draw[thick] (h1.south east) edge (h2.north east);
	\draw[thick] (h1.south west) edge (h2.north west);

	\draw[thick] (h2.south east) edge (h3.north east);
	\draw[thick] (h2.south west) edge (h3.north west);

	\draw[thick] (h3.south east) edge (h4.north east);
	\draw[thick] (h3.south west) edge (h4.north west);

	\draw[thick] (h4.south east) edge (h5.north east -| h4.south east);
	\draw[thick] (h4.south west) edge (h5.north west -| h4.south west);

	\draw[thick] (h5.south east) edge (h6.north east -| h5.south east);
	\draw[thick] (h5.south west) edge (h6.north west -| h5.south west);

\end{tikzpicture}
\caption{Schematic description of our neural network. The weights from~\eqref{eq:reg_mat_dl} are denoted by~$\bm{w} =\{ w_1,\dots,w_{n_m}\}$. The dimensions of $\bm{h}^{(1)}, \bm{h}^{(2)}, \bm{h}^{(3)}$ and $\bm{h}^{(4)}$ are $600,300,200$ and $500$ respectively. Note that the dimension of the final layer is equal to the number of matrices used.}
\label{fig:Neural_network}
\end{figure}
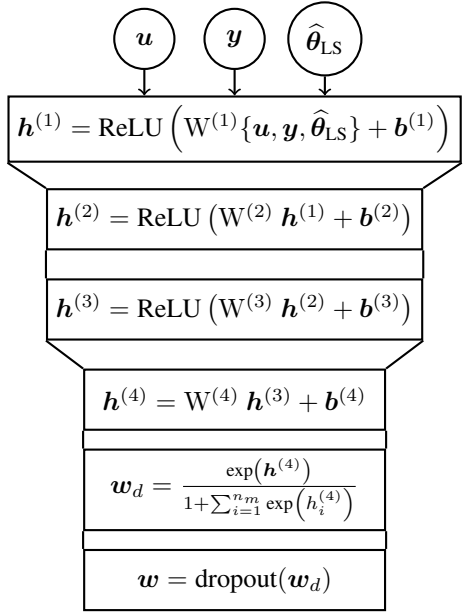

The resulting network is schematically illustrated in Fig.~\ref{fig:Neural_network}.

\subsection{Normalizing the data}
\label{sec:NormalMod}
A key aspect to successfully train neural networks is the normalization of the input and output of the network by subtracting the mean and dividing with the standard deviation. We notice that we can, without loss of generality, normalize each data example of~$\bm{y}$ and~$\bm{u}$ if we at the same time do the corresponding scaling of the impulse response~$\thetab$ to keep the analytic relationship intact.  

The non-regularized least square solution, $\thetabhls$, that we use as an input to the network is also straightforward normalize with the statistics form the true impulse response calculated from the training data. Finally, we want to normalize the networks dependence of the vectors~$\bm{s}_i$. The outer product of these vectors should correspond to the optimal regularization (see Section~\ref{sec:RegMod} and Appendix~\ref{sec:opt_reg}). To enforce that they follow the same statistics, we initialize the vectors~$\bm{s}_i$ with unit Gaussian and then multiply we the standard deviation and add the mean of the true impulse response. 

\begin{algorithm}
\caption{Training procedure}\label{alg:algorithm}
\begin{algorithmic}[1]
	\State Collect evaluation data.
	\State Collect training data with similar behaviour as evaluation data (e.g. through simulations).
	\State Normalize $\bm{u}$ and $\bm{y}$  for each example (Section \ref{sec:NormalMod}).
	\State Normalize $\thetabhls$ with statistics from the whole training dataset  (Section \ref{sec:NormalMod}).
	\State Find $\hat \eta$ using training data by minimizing Equation \eqref{eq:est_sum}.
	\State Use $\hat \eta$ to predict $\thetabh$ on evaluation data.
\end{algorithmic}
\end{algorithm}

\section{Experiment}
\label{sec:exp}
The model explained in Section~\ref{sec:DL} is implemented using Tensorflow \citep{Tensorflow} and our implementation of the model is available on GitHub\footnote{https://github.com/carl-andersson/impest}. 

\subsection{Simulate data using \texttt{rss}}
To train the model, we use an artificial distribution over real systems to produce input and output sequences along with their true impulse responses. The artificial distribution over systems we use is MATLAB's \texttt{rss} function. This function is not ideal as was recently pointed out by for example \citet{Rojas2015}. Hence, there is potential to further improve the results by making use of better data.

To generate data we use the same method as \citet{Chen2012} with some minor alterations concerning the signal to noise ratio (SNR). The full procedure follows as:
\begin{enumerate}
	\item Sample a system of order 30 using a slightly modified version of MATLAB's \texttt{rss} where the probability of having an integrating pole is zero. 
	\item Sample~$N = 125$ timesteps from the system at a sampling rate of 3 times the bandwidth, i.e.,
		\begin{verbatim}
			bw = bandwidth(m);
			f = 3*(bw*2*pi); 
			md = c2d(m,1/f,'zoh');
    	\end{verbatim}
   	
   	\item Calculate the true FIR parameters as the truncated impulse response to length~$n=50$.
   	\item Sample white noise as an input sequence~$\bm{u}$ and get the corresponding output sequence~$\y^*$. Add noise to the output with the SNR drawn randomly from a uniform distribution between 1 and 10, i.e., the noise added has variance that is between 1 and 1/10 of the variance in the output sequence. 
   	\item Repeat all steps until we have~$M$ examples from~$M$ different systems.
\end{enumerate}

These~$M$ training examples are then used in Equation~\eqref{eq:est_sum} and are called the training data. Using the same method we generate a validation set which we also split in two sets depending on the SNR, one with SNR larger than~$5.5$ and one set with SNR less than~$5.5$ with roughly $M_v\approx 5\thinspace000$ examples in each.

\subsection{Evaluation metrics}
To evaluate the performance of the model we use a metric that is calculated as the mean squared error of the estimate  normalized with the mean squared error of the least squares solution without any regularization. We denote this metric by~$S$, i.e.,
\begin{align}
\label{eq:score}
	S = \frac{1}{M_v} \sum_{i=1}^{M_v} \left [ \frac{\left \lVert \thetabh \left (f_\text{DL}(\bm{u}^{(i)},\bm{y}^{(i)};\eta) \right ) - \thetab^{(i)}_0 \right \rVert^2}{\left \lVert \thetabhls^{(i)}-\thetab^{(i)}_0  \right \rVert^2} \right ],
\end{align}
where $M_v$ denotes the number of examples in the validation set. This metric makes sure that each impulse response gets equal weighting when computing the performance of the algorithm and measures the average effect of the regularization. A perfect match for this measure corresponds to a measure of $0$. 

\citet{Chen2012} make use of a slightly different metric defined as
\begin{align*}
	\widetilde{S} = \frac{1}{M_v} \sum_{i=1}^{M_v} 100\left ( 1 - \left [ \frac{\left \lVert  \thetabh \left (f_\text{DL}(\bm{u}^{(i)},\bm{y}^{(i)};\eta) \right ) - \thetab^{(i)}_0  \right \rVert^2 }{\left \lVert \thetab_0^{(i)} - \bar{\thetab}_0^{(i)}   \right \rVert^2} \right ] \right )
\end{align*}
where $\bar{\thetab}_0^{(i)}$ is  the mean of $\thetab_0^{(i)}$. This metric averages over a so-called 'model fit', i.e. how well the estimated parameters fit the true impulse response. Besides the shifting and scaling in~$\widetilde{S}$, the only difference between the two metrics is the normalization factor used, where~$S$ is normalized with the least squares estimation error and~$\widetilde{S}$ is normalized with the variance of the true impulse response. We have empirically observed that the terms in~$\widetilde{S}$ might vary a lot between different examples and the average might thus be dominated by a few examples leading to measures that are hard to interpret. Our slightly modified metric~$S$ will on the other hand measure the average effect of using a regularization method compared to not using a regularization method, which seem to result in a more stable performance indicator.

\subsection{Simulation results}
\label{sec:simulation}
The model is trained using $M = 1\thinspace000\thinspace000$ training examples for roughly 2.5 hours using a desktop computer with an Nvidia Titan Xp GPU. The chosen hyperparameters of the model is described in Figure \ref{fig:Neural_network}. Note that while training require a GPU with large memory, the evaluation can easily be done in CPU on an ordinary computer. We are using early stopping as a stopping criteria even though the model essentially does not seem to overfit with this amount of training data. The model is not very dependent on the number of examples in the training data either. Even with $M=10\thinspace000$ it managed to achieve comparable results to previous methods.

\begin{figure}[t]
	\includegraphics[width=1.0 \linewidth]{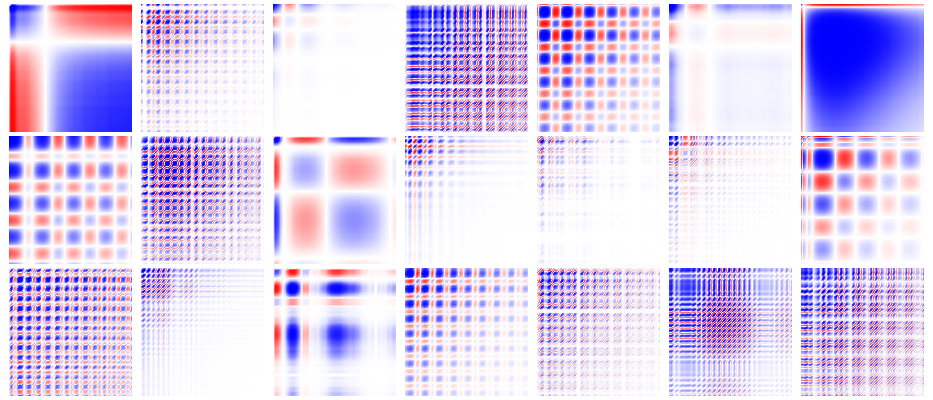}
	\captionof{figure}{Illustration of 21 matrices from $\bm{s}_i {\bm{s}_i}^T$ after training. The matrices are rescaled to the interval $[+1,-1]$, where blue indicates a positive value, white around zero and red indicates a negative value. The upper left corner corresponds to the lowest index of the estimate~$\thetab$.}
	\label{fig:training_matrices}
\end{figure}

\begin{figure}
	\hfill
	\begin{subfigure}[t]{0.3\linewidth}
		\includegraphics[width=\linewidth]{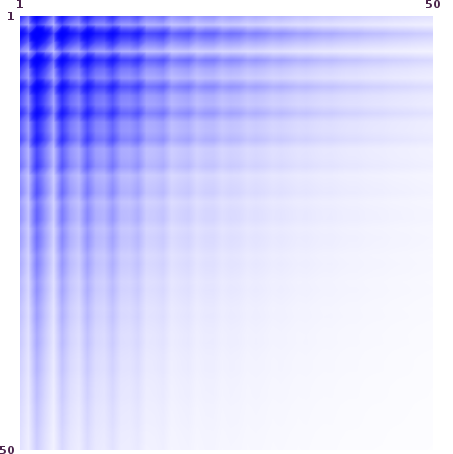}
		\centering
		\caption{Optimal regularization}
	\end{subfigure}
	\hfill
	\begin{subfigure}[t]{0.3\linewidth}
		\includegraphics[width=\linewidth]{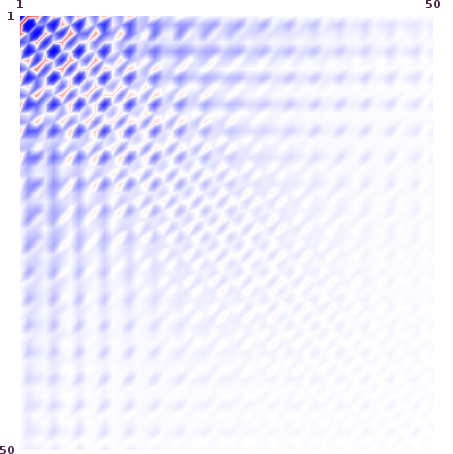}
		\centering
		\caption{Deep learning regularization}
	\end{subfigure}
	\hfill
	\begin{subfigure}[t]{0.3\linewidth}
		\includegraphics[width=\linewidth]{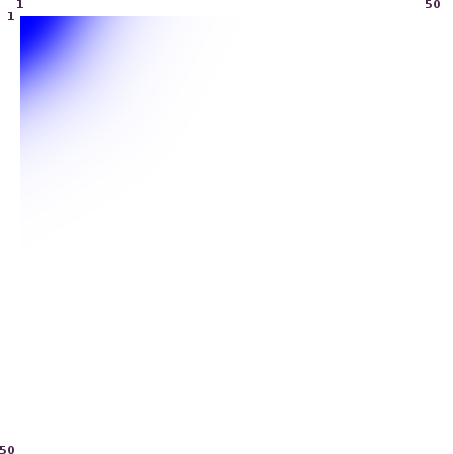}
		\centering
		\caption{Gaussian process regularization}
	\end{subfigure}
	\hfill
	\centering
	\caption{Comparison between rescaled inverse regularization matrices for a validation example where our method captures the behaviour of the optimal regularization matrix. }
	\label{fig:good_ex}
\end{figure}

\begin{figure}
 \centering
 \begin{tikzpicture}
    \definecolor{color4}{HTML}{2084C5}
    \definecolor{color2}{HTML}{EDB120}
    \definecolor{color3}{HTML}{D95319}
    \definecolor{color1}{HTML}{7E2F8E}
    \pgfplotsset{
     tick label style = {font=\small},
	 every axis label = {font=\small},
  	 legend style = {font=\small},
	 label style = {font=\small}
    }
 	\pgfsetlinewidth{20};
     \begin{axis}[
         height = 0.7 \linewidth,
         width = \linewidth,                    
         xlabel= {Coefficient index},
         ylabel= {Coefficient value},
         xmin=0,xmax=50,
         grid = major ,
         height = 5cm,
       ]  
	       
       \addplot+[mark=none,thick,color1]  table[x index=0,y index=1] {data/estimates.dat};
       \addlegendentry{\small{Gaussian process}};
	   \addplot+[mark=none,thick,color2]  table[x index=0,y index=2] {data/estimates.dat};
	   \addlegendentry{\small{Deep learning}};
	   \addplot+[mark=none,thick,color3]  table[x index=0,y index=3] {data/estimates.dat};
       \addlegendentry{\small{Least square}};           
	   \addplot+[mark=none,thick,color4]  table[x index=0,y index=4] {data/estimates.dat};
       \addlegendentry{\small{True}};   
    \end{axis}
 \end{tikzpicture}

	\caption{Estimates of the impulse response coefficients,~$\thetabh$, using the inverse regularization matrices from Fig.~\ref{fig:good_ex} and the same input and output sequence.}
	\label{fig:Estimates}
\end{figure}
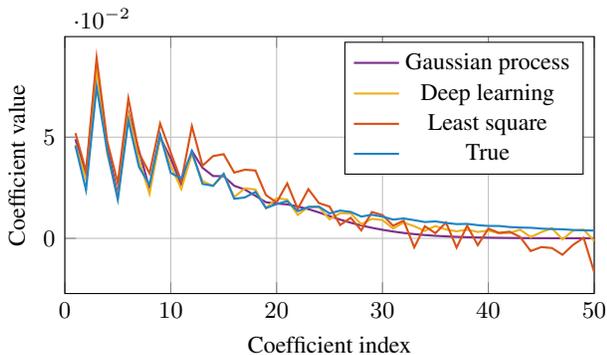

Fig.~\ref{fig:training_matrices} shows a subset of the matrices $\bm{s}_i\bm{s}_i^T$ from~\eqref{eq:reg_mat_dl} after training. Note that the matrices have an oscillating pattern with different periods and a decay towards zero for parameters with high index (lower right corner). Fig.~\ref{fig:good_ex} shows three different regularization matrices for an example in the validation dataset. We can see that the deep learning regularization seems to capture the behaviour of the optimal regularization matrix fairly well. In Fig.~\ref{fig:Estimates} we can compare the estimates from the different regularization approaches.  

The trained model does not produce a useful regularization matrix for all examples. In cases where it fails the neural network seems to fail in the same way for all examples by producing a similar regularization matrix for each example with a bad performance for~$S$ as consequence. Despite this problem, the model manages to produce average results which are comparable or better than previous methods which is reflected in Table~\ref{tab:res}. 
We can see that the performance decreases when the SNR gets larger. This is due to the improved effectiveness of the least squares estimate and we are thus less dependent on the regularization. For comparison we also present the result for the optimal regularization which of course is unachievable since it depends on the true impulse response, but it is still an useful lower bound.
\begin{figure}[h]
	\begin{minipage}{0.9 \columnwidth}
		\captionof{table}{Comparing the different models using the metric from~\eqref{eq:score} evaluated using the validation set. LS stands for Least Squares, OR stands for Optimal regularization (see Appendix \ref{sec:opt_reg}), GP stands for Gaussian process regularization and DL stands for deep learning regularization. LS, OR and GP are not data driven approaches and they are thus not dependent on any training data.}
		\centering
		\begin{tabular}{  c |  c |  c }
			\textbf{Model} & \quad $\text{SNR} < 5.5$ \quad & \quad $\text{SNR} > 5.5$ \quad \\
			\hline
			LS & 1       &  1 \\  
			OR & 0.04    & 0.05\\
			GP & 0.31    & 0.40 \\
			DL & 0.20    & 0.23 \\
			\multicolumn{3}{c}{}
		\end{tabular}
		\label{tab:res}
	\end{minipage}
\end{figure}

\subsection{Real data results}
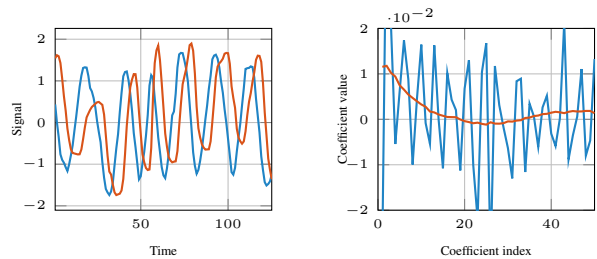
\begin{figure}
	\hfill
	\begin{subfigure}[t]{0.45\linewidth}
		\centering
 \begin{tikzpicture}
    \definecolor{color4}{HTML}{2084C5}
    \definecolor{color2}{HTML}{EDB120}
    \definecolor{color3}{HTML}{D95319}
    \definecolor{color1}{HTML}{7E2F8E}
    \pgfplotsset{
     tick label style = {font=\tiny},
	 every axis label = {font=\tiny},
  	 legend style = {font=\tiny},
	 label style = {font=\tiny}
    }
 	\pgfsetlinewidth{20};
     \begin{axis}[
         height = 0.7 \linewidth,
		 width = 1.15\linewidth,                     
         xlabel= {Time},
         ylabel= {Signal},
         y label style={at={(axis description cs:-0.1,0.5)}},
         xmin=1,xmax=125,
         height = 4cm,
         grid = major ,
       ]  
	       
       \addplot+[mark=none,thick,color4]  table[x index=0,y index=1] {data/datainout.dat};
	   \addplot+[mark=none,thick,color3]  table[x index=0,y index=2] {data/datainout.dat};
    \end{axis}
 \end{tikzpicture}
	\caption{An input sequence (blue) and the corresponding output (red) for a real system.}
	\label{fig:data}
	\end{subfigure}
	\hfill
	\begin{subfigure}[t]{0.45\linewidth}
		\centering
 \begin{tikzpicture}
    \definecolor{color4}{HTML}{2084C5}
    \definecolor{color2}{HTML}{EDB120}
    \definecolor{color3}{HTML}{D95319}
    \definecolor{color1}{HTML}{7E2F8E}
    \pgfplotsset{
     tick label style = {font=\tiny},
	 every axis label = {font=\tiny},
  	 legend style = {font=\tiny},
	 label style = {font=\tiny}
    }
 	\pgfsetlinewidth{20};
     \begin{axis}[
         height = 0.7 \linewidth,
         width = 1.15\linewidth,                    
         xlabel= {Coefficient index},
         ylabel= {Coefficient value},
         y label style={at={(axis description cs:-0.1,0.5)}},
         xmin=0,xmax=50,
         ymin=-0.02,ymax=0.02,
         height = 4cm,
         grid = major ,
       ]  
	   
	   \addplot+[mark=none,thick,color4]  table[x index=0,y index=2] {data/dataest.dat};
	       
       \addplot+[mark=none,thick,color3]  table[x index=0,y index=1] {data/dataest.dat};

    \end{axis}
 \end{tikzpicture}
	\caption{Impulse response estimates using the data in Fig.~\ref{fig:data} for regular least squares (blue) and our method (red).}
	\end{subfigure}
	\hfill
	\centering
	\caption{Our method applied to an example from a real system and real measured data.}
	\label{fig:realdata}
\end{figure}

To show that our method at least does not give unreasonable estimates for real systems and data we test our method on data measured at a processing plant. See e.g.~\cite{BittencourtIPF:2015} for an introduction to this problem area.  We do not know the true parameter values for these systems, implying that we cannot evaluate the performance of the estimates. We use the same trained network as we evaluated in the previous section. In Fig.~\ref{fig:realdata} we simply present an input sequence and the corresponding output sequence from a real system together with the estimates produced by our method compared to using least squares without regularization. We note also that the results seems reasonable and that the regularization removes a lot of noise in the estimation.

\section{Conclusion}
In this paper we present a method to regularize an impulse response estimation problem. We train a model with simulated data in a data-driven fashion to encode this regularization with parameters in a neural network. A trained model can then be used to improve the mean squared error of the new estimations. 
The results of our method seems promising and there is plenty of scope for future work along this line of research, both when it comes to impulse response estimation, but also for other problems. We find it especially interesting that this model can mimic the optimal regularization matrix to higher degree than previous methods which we believe is the reason for why it sometimes produce better estimates.

Although training our model is quite time consuming, estimating the impulse response using our method is very fast since it only involves a couple of matrix  multiplications to compute the regularization matrix, whereas the method of \citet{Chen2012} needs to solve an optimization problem for each example. 

\section{Future work}
We are planning to further investigate how one can make use of real data from e.g. the process industry. This would make it possible to use large amounts of collected data to improve the estimated parameters in a data-driven manner. The process industry has a lot of data available and makes extensive use of linear models.

The idea of learning a prior by representing it with a regularization matrix in the form of a neural network is not unique to the problem of estimating the impulse response. It could easily be generalized to other situations where the least squares solution is available but the prior of the solution is either unknown or intractable. If one can simulate many such systems at low cost, or have data from true systems available, formulating a regularization matrix as a neural network might be a tractable way of regularizing the estimate.

The presented approach can easily be extended to multi-input multi-output systems where the only difference is that the dimension of the input and output sequences and the parameters~$\thetab$ increases. The deep structure of the model automatically induces any relevant connection between the different system input and output components present in the training data. 

Finally we want to stress that this is just one example of what one might do with deep learning in system identification. There might and should be other areas where it is possible to make use of either simulated or real data to improve standard methods, or invent new methods for system identification. For example it might be worth looking into Recurrent Neural Networks (RNN) such as Long Short Term Memory (LSTM)\citep{Hochreiter1997}, Gated Recurrent Units (GRU) \citep{Cho2014} or Stocastic Recurrent Neural Network (SRNN)\citep{Fraccaro2016} and apply it in a system identification setting or even to bring some of the system identification knowledge of dynamical systems to the field of deep learning. 

\section*{Acknowledgements}
We would like to thank Professor Krister Forsman at Perstorp and the Norwegian University of Science and Technology for providing us with the real world data. We would also like to thank Dr Jack Umenberger for useful feedback on an early draft of this work.

\begin{small}
	\bibliographystyle{abbrvnat}
	\bibliography{refs}             
\end{small}

\appendix                                                 
\section{Optimal regularization}
\label{sec:opt_reg}
The optimal regularization matrix is a term coined by \citet{Chen2012}. It corresponds to the regularization matrix that is optimal in the sense that it minimizes the mean squared error. The optimal regularization matrix can be written as, $P = \sigma^{-2} \thetab_0 \thetab_0^\T$ \citep{Chen2012}, where $\sigma^2$ is the variance of the additive noise~$v(t)$ in~\eqref{eq:system}, $\thetab_0$ is the true impulse response of the system without noise.

\end{document}